\documentclass[onecolumn,floatfix,superscriptaddress,showpacs,showkeys,nofootinbib]{revtex4}%
\usepackage{epsfig}
\usepackage{amssymb,latexsym,amsmath}
\newcommand{\eq}[1]{\begin{align} #1 \end{align}}

\usepackage{bm}
\usepackage{amsmath}
\usepackage{empheq}
\usepackage{graphicx}
\usepackage{subfigure}
\usepackage{enumerate}
\usepackage{soul}
\usepackage[colorlinks=true,linktocpage=true,linkcolor=blue,citecolor=blue,allcolors=blue]{hyperref}
\usepackage[usenames,dvipsnames]{color}
\usepackage{url}

\begin{document}

\title{Multi moment cancellation of participant fluctuations - MMCP method}
\author{Viktor Begun} \email{viktor.begun@gmail.com}
\author{Maja Mackowiak-Pawlowska} \email{majam@if.pw.edu.pl}

\affiliation{Warsaw University of Technology, Faculty of Physics, Koszykowa 75, 00-662 Warsaw Poland}

\begin{abstract}
We propose a new way to correct for finite centrality bin width effect i.e. participant fluctuations in fluctuation analysis in high energy nucleus-nucleus collisions.
The MMCP method allows to separate participant fluctuations and obtain fluctuations from one participant - a source - from a combination of the experimentally measured first four moments.
The EPOS model is used for the numerical check of the MMCP for the net electric charge fluctuations in the forward rapidity region in Ar+Sc reactions at beam momentum 150~GeV/c.
We show that using the existing methods - decreasing a centrality bin width, or using the Centrality Bin Width Correction procedure, one may still leave some residual participant fluctuations in the sample.
Moreover, we show that the Centrality Bin Width Correction procedure may alter the fluctuation measures.
The most important advantage of the MMCP is it's precision even when the amount of measured events does not allow to decrease the centrality bin width, or the experimental determination of participants is difficult, e.g. in collider experiments.
Even for the largest centrality bin in the considered case, $c=0-20\%$, the relative error of the MMCP for the scaled variance of a source is below 2\%.
It is especially important in determination of the base line of the fluctuations in the search for the QCD Critical Point and the signals of the QCD phase transition.

\end{abstract}



\date{\today}

\maketitle

%
\section{Introduction}\label{sec-Intro}
%
%
The measurements of multiplicity fluctuations belong to the milestones of the heavy ion studies at SPS~\cite{Gazdzicki:995681,Grebieszkow:2009jr,Grebieszkow:2016cza} and RHIC~\cite{Aggarwal:2010wy,Adare:2015aqk,Thader:2016gpa,Luo:2015doi,Luo:2017faz}. Their aim is to find the QCD critical point, and study its properties. The only experimentally controllable way to probe the phase diagram of strongly interacting matter is by studying interactions of different system size nuclei at various energies. One of the main background effects in such study is the fluctuations of nucleon participants, $N_{\rm P}$. It is the number of nucleons that interacted inelastically and produced other particles during nucleus-nucleus (A+A) collision. This number fluctuates from event to event, reflecting the geometry of the collision, and, possibly, hiding fluctuations from other sources.
There are several popular ways of addressing participant fluctuations:
\begin{enumerate}[(i)]
	\item \label{centbin} the selection of as narrow centrality bins as possible,
	\item \label{cbwc} the Centrality Bin Width Correction procedure (CBWC)~\cite{Luo:2011ts},
	\item \label{siq} the use of strongly intensive measures (SIM)~\cite{Gazdzicki:1992ri,Gorenstein:2011vq,Poberezhnyuk:2015dta,Sangaline:2015bma,Broniowski:2017tjq},
\end{enumerate}
see also for example~\cite{Altsybeev:2016uql}. 
The first two methods are based on the effect that participant fluctuations decrease with decreasing centrality bin width. The constrain of participant fluctuations via centrality selection (\ref{centbin}) in fixed target experiments is more precise than in collider experiments, a, as one can put a detector at the beam line and measure the number of projectile spectators. This is the number of protons and neutrons from the projectile nuclei that did not participate in the collision. If the detector measures zero signal, then it was the 'head on' collision - almost all protons and neutrons were participants.
In collider experiments it is impossible to constrain participant fluctuations with such precision. The projectile spectators fly to the beam pipe of a collider together with the nuclei that did not interact.
Another source of participant fluctuations are target participants~\cite{Gazdzicki:2005rr}. These remain unconstrained usually both in fixed target and collider experiments, although for central collisions such fluctuations should be small.
However, even small participant fluctuations may affect fluctuations described by higher order moments.
Therefore, a reliable way to minimize participant fluctuations is necessary.

The idea behind the CBWC (\ref{cbwc}) is that the centrality bin is divided into as narrow sub-bins as possible, and then the fluctuations in the bin are calculated from the fluctuations in the sub-bins. The division is proposed to suppress participant fluctuations, while the recalculation is needed in order to decrease statistical uncertainty due to small number of events in each sub-bin. The CBWC was used in STAR and ALICE Collaborations~\cite{Luo:2011ts,Adamczyk:2013dal,Adamczyk:2014fia,Luo:2015ewa,Chen:2016xyu,Mukherjee:2016hrj}.
Neither the decrease of the centrality, nor the CBWC can guarantee the full removal of participant fluctuations. 
%

Strongly intensive measures (\ref{siq}) are independent of system's volume and it's event-by-event fluctuations. 
In order to calculate such variables, two quantities describing an event are required. Then, a special combination of them cancels volume fluctuations under assumptions that both quantities are produced in the same volume and with the same volume fluctuations in grand-canonical ensemble. The same result can be obtained in a wounded nucleon model~\cite{Bialas:1976ed}, or in the Independent Particle Production model (IPM)~\cite{Begun:2016sop}. 

We propose a different approach - to cancel participant fluctuations in a combination of several high fluctuation moments of the same quantity, e.g. particle type.
We call our new method - multi moment cancellation of participant fluctuations (MMCP).
It exploits the relations between the measured first four moments of the multiplicity distribution in IPM. We test how participant fluctuations depend on the centrality bin width (\ref{centbin}), and compare MMCP to CBWC~(\ref{cbwc}). The new method does not require the basic assumption of SIM that volume and volume fluctuations are the same for two quantities, because it can be necessary to avoid it. For example, pions come mainly from resonance decays, while heavy multi-strange particles are produced directly in an A+A reaction. The assumption of the same volume fluctuations may be invalid for them. Then combining pions and multi-strange particles into SIM may lead to a false conclusion. The most commonly used particles for SIM are pions and kaons. They both have a large contribution of resonances, therefore, the usage of SIM is acceptable at most energies  for them. However, resonances are not enough to explain pions and kaons at all energies. For example, the $K^+/\pi^+$ ratio exhibits the Horn structure at about $30A$~GeV energy, while $K^-/\pi^-$ does not, see~\cite{Gazdzicki:1998vd} and, e.g.~\cite{Mackowiak-Pawlowska:2017rcx}. Therefore, SIM may be invalid in the situations, where the SIM are the most needed. The MMCP may give an important advantage in these cases.
%

%
As the test we study the net electric charge fluctuations in Ar+Sc reactions generated in EPOS 1.99 model~\cite{Werner:2008zza,Pierog:2009zt} at forward rapidities and beam momentum $p_{beam}=150$~GeV/c. The Ar+Sc reactions are selected, because they are studied as a part of NA61/SHINE energy and system size scan program~\cite{Gazdzicki:995681}. EPOS is used, because it is a realistic model of A+A collisions, which allows to keep track of the number of participants, and calculate their contribution explicitly.

The paper is organized as follows. In Section~\ref{sec-Part} we introduce the necessary relations for our studies from the IPM. Section~\ref{sec-EPOS} contains the results of the simulations of Ar+Sc reactions in EPOS and the tests of the MMCP. In Section~\ref{sec-Comp} we compare MMCP and CBWC methods.  Section~\ref{sec-Concl} concludes the paper.

%
\section{Participant fluctuations}\label{sec-Part}
%
%
A multiplicity distribution, $P(N)$, can be characterized by central moments, $\mu_n$, which are defined as follows
 \eq{\label{ctr-mom}
 \mu_n ~=~ \sum_N\left(N-\langle N\rangle\right)^n P(N)  ~=~ \langle \left(N-\langle N\rangle\right)^{n}\rangle~,
 }
where
 \eq{
 \langle N^n\rangle ~=~ \sum_N N^n\, P(N)~
 }
are raw moments. They are related to cumulants,
 \eq{
 \kappa_2~=~\mu_2~,&& \kappa_3~=~\mu_3~,&& \kappa_4~=~\mu_4~-~3\mu_2^2~,&&\ldots,
 }
%
%
The frequently used cumulant ratios - scaled variance, normalized skewness and normalized kurtosis - are:
 \eq{\label{w-Ss-Ks2}
 \omega~=~\frac{\kappa_2}{\langle N\rangle} ~=~ \frac{\sigma^2}{\langle N\rangle} ~, &&
 S\,\sigma~=~\frac{\kappa_3}{\kappa_2}~, &&
 \kappa\,\sigma^2 ~=~ \frac{\kappa_4}{\kappa_2}~,
 }
where $\sigma$ is standard deviation. The quantities in Eq.~(\ref{w-Ss-Ks2}) are intensive in an IPM, i.e. they do not depend on average number of participants. However, they are not strongly intensive, because they depend on participant fluctuations, i.e. higher order moments of participant distribution. To assure clarity of the subsequent consideration, the following notations are applied:
\begin{itemize}
	\item the index $\rm P$ denotes quantities obtained from  {\it the participant distribution},
	\item the index $\rm A$ denotes quantities obtained from the distribution for a single participant - {\it a source}\footnote{An existence of such a distribution is assumed, and then it`s characteristics are obtained using the information about the participant distribution. A source defined in this way represents a combination of all sources that produce the quantity of interest. For example, if net charge is considered, then a source is anything that produces net charge and it`s fluctuations in A+A reactions.},
	\item the {\it net charge} or lack of indices indicates quantities which are calculated from net electric charge distribution created in Ar+Sc reaction with selected centrality, containing both the fluctuations coming from a source and participant fluctuations.
\end{itemize}
An IPM assumes that the number of particles of interest $N$ is given by the sum of contributions $n_i$ from $N_{\rm P}$ participants,
\eq{
 N~=~\sum_{i=1}^{N_{\rm P}}n_i~,
}
which are {\it identical} on average,
 \eq{
 \langle n_i\rangle ~=~ \langle n_j\rangle ~=~ \langle n_{\rm A}\rangle~.
 }
The average multiplicity of produced particles $\langle N\rangle$ is then proportional to the average number of particles from one source, $\langle n_{\rm A}\rangle$, and to the average number of participants $\langle N_{\rm P}\rangle$,
 \eq{\label{Nav}
 \langle N\rangle ~=~ \langle n_{\rm A}\rangle\, \langle N_{\rm P}\rangle~.
 }
The additional assumption that particles from different sources are {\it independent}, 
 \eq{
 \langle n_i\,n_j\ldots n_k\rangle ~=~ \langle n_i\rangle \langle n_j\rangle\ldots\langle n_k\rangle ~=~ \langle n_{\rm A}\rangle^k~,
 } 
allows to obtain scaled variance,
 \eq{\label{w}
 \omega 
~=~ \omega_{\rm A} ~+~ \langle n_{\rm A}\rangle ~ \omega_{\rm P}~,
 }
normalized skewness, 
 \eq{\label{Ssig}
 S\,\sigma ~=~ \frac{\omega_{\rm A}~S_{\rm A}\,\sigma_{\rm A}
 ~+~ \langle n_{\rm A}\rangle~\omega_{\rm P}\left[~3\,\omega_{\rm A} ~+~ \langle n_{\rm A}\rangle\,S_{\rm P}\,\sigma_{\rm P}~\right] }
        {\omega_{\rm A} ~+~ \langle n_{\rm A}\rangle ~ \omega_{\rm P}}~,
 }
normalized kurtosis,
 \eq{\label{ks2}
 \kappa\,\sigma^2
 ~=~ \frac{\omega_{\rm A}~\kappa_{\rm A}\,\sigma_{\rm A}^2
        ~+~\langle n_{\rm A}\rangle~\omega_{\rm P}~ \left[~\langle n_{\rm A}\rangle^2~\kappa_{\rm P}\,\sigma_{\rm P}^2
        ~+~\omega_{\rm A}\left(~
           3\,\omega_{\rm A}
        ~+~4\,S_{\rm A}\,\sigma_{\rm A}
        ~+~6\,\langle n_{\rm A}\rangle\,S_{\rm P}\,\sigma_{\rm P}
           ~\right)~\right]}
     {\omega_{\rm A} ~+~ \langle n_{\rm A}\rangle ~ \omega_{\rm P}}~,
 }
and any other combination of higher moments\footnote{See also the derivation based on the assumption of the existence of the cumulant generating function with separable volume~\cite{Sangaline:2015bma,Skokov:2012ds,Braun-Munzinger:2016yjz,Xu:2016skm,Broniowski:2017tjq}}~\cite{Begun:2016sop}.

The values of interest are the fluctuations from a source: $\omega_{\rm A}$, $S_{\rm A}\,\sigma_{\rm A}$, $\kappa_{\rm A}\,\sigma_{\rm A}^2$. They are mixed with the fluctuations of the participants $\omega_{\rm P}$, $S_{\rm P}\,\sigma_{\rm P}$, $\kappa_{\rm P}\,\sigma_{\rm P}^2$. Moreover, higher order fluctuation moments have the contribution from lower order moments,
 \eq{\label{Nn0}
\mu_n ~=~ {\mathcal{F}}\left(\langle N^1 \rangle, \langle N^2 \rangle, \ldots \langle N^n \rangle \right)~,
 }  
where  $\mathcal{F}$ just denotes that the value on the l.h.s. of Eq.~(\ref{Nn0}) is a function of the values in the brackets on the r.h.s. of Eq.~(\ref{Nn0}).  Equation (\ref{Nn0}) is the usual property of a central moment, which can be seen by expanding the binomial in Eq.~(\ref{ctr-mom}). The assumption that there are participant or volume fluctuations and the fluctuations from a source makes $\mu_n$ dependent on the moments of both, a source and participants,
 \eq{\label{Nn}
 \mu_n ~=~ {\mathcal{F}}\left( \langle n_{\rm A}^1 \rangle, \langle n_{\rm A}^2 \rangle, \ldots \langle n_{\rm A}^n \rangle,\langle N_{\rm P}^1 \rangle, \langle N_{\rm P}^2 \rangle, \ldots \langle N_{\rm P}^n \rangle \right)~.
 }  
%
It gives $n$ measures versus $2n$ unknowns for their description. However, it is {\it unavoidable} situation, if one is not sure that the system consists of only one type of fluctuations that are directly represented by the fluctuation moments as in~(\ref{Nn0}). 

Methods mentioned in the introduction address this obstacle to a various degree. The method (\ref{centbin}) may still leave some unknown fraction of participant fluctuations even for a very narrow centrality bin. The method (\ref{cbwc}) was invented because of the design features of the STAR detector. It reduces {\it statistical} fluctuations only, leaving some participant fluctuations~\cite{Mohanty:priv}, see next section. The introduction of strongly intensive measures (\ref{siq}) requires two types of measured values, i.e. multiplicities of particle type A, $N_{\rm A}$, and the multiplicity of particle type B, $N_{\rm B}$, e.g. pions and kaons, and the assumption that all corresponding participant fluctuations moments are the same $\langle N_{\rm P_A}^n\rangle=\langle N_{\rm P_B}^n\rangle=\langle N_{\rm P}^n\rangle$. This may not be true, because participant fluctuations may influence one particle type more than another.

We propose another method - to find a specific combination of several high moments, which cancel participant fluctuations in low moments. We face the same difficulty as in Eq.~(\ref{Nn}): n measures and 2n unknowns. However, it is possible to overcome it assuming that participant fluctuations are small, and neglect some participant moments.

First of all, let us assume that the methods (\ref{centbin}) and (\ref{cbwc}) were effective enough to make scaled variance for the fluctuations from a source close to the measured fluctuations,
 \eq{\label{alpha}
 \omega~\simeq~\omega_{\rm A}~,&&\text{then}&&\alpha~=~\frac{\omega-\omega_{\rm A}}{\omega_{\rm A}}~=~\langle n_{\rm A}\rangle~\frac{\omega_{\rm P}}{\omega_{\rm A}}~\ll~1~.
 }
The agreement between $\omega$ and its strongly intensive analog $\Omega$ presented in~\cite{Seryakov:2017sss} for Ar+Sc reactions in $0-0.2\%$ centrality bin justifies this assumption. In the next section we show by direct calculation that the assumption (\ref{alpha}) is valid in Ar+Sc even for centralities up to $0-20\%$ in EPOS 1.99 model.
%
One can rewrite Eqs.~(\ref{Ssig}) and (\ref{ks2}) using $\alpha$ parameter (\ref{alpha}), and expand them in Taylor series, leaving only the terms that are proportional to zero-th and to the first order of $\alpha$: 
 \eq{\label{Ssig-1}
 S\,\sigma &~=~ \frac{S_{\rm A}\,\sigma_{\rm A}
 ~+~ \alpha\left[~3\,\omega_{\rm A} ~+~ \langle n_{\rm A}\rangle\,S_{\rm P}\,\sigma_{\rm P}~\right] }
        {1 ~+~ \alpha}
  ~\simeq~ S_{\rm A}\,\sigma_{\rm A}\,(1-\alpha) ~+~ \alpha\left[~3\,\omega_{\rm A} ~+~ \langle n_{\rm A}\rangle\,S_{\rm P}\,\sigma_{\rm P}~\right]~,&& \alpha~\ll~1~,
 }
and
 \eq{\nonumber
 \kappa\,\sigma^2
 &~=~ \frac{\kappa_{\rm A}\,\sigma_{\rm A}^2
        ~+~\alpha~ \left[~\langle n_{\rm A}\rangle^2~\kappa_{\rm P}\,\sigma_{\rm P}^2
        ~+~\omega_{\rm A}\left(~
           3\,\omega_{\rm A}
        ~+~4\,S_{\rm A}\,\sigma_{\rm A}
        ~+~6\,\langle n_{\rm A}\rangle\,S_{\rm P}\,\sigma_{\rm P}
           ~\right)~\right]}
     {1 ~+~ \alpha}
       \\ \label{ks2-1}
 &~\simeq~ \kappa_{\rm A}\,\sigma_{\rm A}^2\,(1-\alpha)
        ~+~\alpha~ \left[~\langle n_{\rm A}\rangle^2~\kappa_{\rm P}\,\sigma_{\rm P}^2
        ~+~\omega_{\rm A}\left(~
           3\,\omega_{\rm A}
        ~+~4\,S_{\rm A}\,\sigma_{\rm A}
        ~+~6\,\langle n_{\rm A}\rangle\,S_{\rm P}\,\sigma_{\rm P}
           ~\right)~\right]~, 
 && \alpha~\ll~1~.
 }

Second, the scaled variance of a source, $\omega_{\rm A}$, competes with the normalized skewness and kurtosis times the number of particles from one source, $ \langle n_{\rm A} \rangle\,S_{\rm P}\,\sigma_{\rm P}$, and $\langle n_{\rm A}\rangle^2\,\kappa_{\rm P}\,\sigma_{\rm P}^2$ in Eqs.~(\ref{Ssig-1}), (\ref{ks2-1}). We assume that their ratio is also small,
 \eq{\label{beta-gamma}
 |\beta| ~=~  \langle n_{\rm A} \rangle~\frac{|S_{\rm P}\,\sigma_{\rm P}|}{\omega_{\rm A}} ~\ll~ 1~, 
 &&
 |\gamma|~=~\langle n_{\rm A}\rangle^2~\frac{|\kappa_{\rm P}\,\sigma_{\rm P}^2|}{\omega_{\rm A}^2}~\ll~1~,
 } 
then 
\begin{empheq}[left = \empheqlbrace]{align}
 \label{wIPM}
 \omega &~=~ \omega_{\rm A}~(1+\alpha)~,
 \\ \label{SsigIPM}
  S\,\sigma 
  &~\simeq~ S_{\rm A}\,\sigma_{\rm A}~(1-\alpha) ~+~ 3\,\alpha\,\omega_{\rm A}~,
  \\\label{ks2IPM}
  \kappa\,\sigma^2
   &~\simeq~ \kappa_{\rm A}\,\sigma_{\rm A}^2~(1-\alpha)
        ~+~3\,\alpha\,\omega_{\rm A}^2~\left[ ~1~+~\frac{4}{3}~\frac{S_{\rm A}\,\sigma_{\rm A}}{\omega_{\rm A}}~\right]~, 
 && \alpha,~|\beta|,~|\gamma|~\ll~1~,
\end{empheq}
where we omitted the terms proportional to $\alpha\,\beta$ and $\alpha\,\gamma$.
%
The assumptions (\ref{alpha}), (\ref{beta-gamma}) is the mathematical expression of the phrase `small participant fluctuations', because one can rewrite the conditions $\alpha\ll1$, $|\beta|\ll1$, $|\gamma|\ll1$ as follows:
 \eq{\label{abc}
 \omega_{\rm P}~\ll~\frac{\omega_{\rm A}}{\langle n_{\rm A}\rangle}~, &&
  |S_{\rm P}\,\sigma_{\rm P}|~\ll~  \frac{\omega_{\rm A}}{\langle n_{\rm A} \rangle}~, &&
  |\kappa_{\rm P}\,\sigma_{\rm P}^2| ~\ll~ \frac{\omega_{\rm A}^2}{\langle n_{\rm A}\rangle^2}~.
 }
Note that small participant fluctuations mean that the {\it ratio} of the scaled variance to the number of particles from one source is large. 
The set of equations (\ref{wIPM}-\ref{ks2IPM}) is underdetermined, because there are three measured values, $\omega$, $S\,\sigma$, $\kappa\,\sigma^2$, and four unknowns $\alpha$, $\omega_{\rm A}$, $S_{\rm A}\,\sigma_{\rm A}$, and $\kappa_{\rm A}\,\sigma_{\rm A}^2$. Therefore, one can not solve  (\ref{wIPM}-\ref{ks2IPM}), but it is possible to express $\omega_{\rm A}$, $S_{\rm A}\,\sigma_{\rm A}$, and $\kappa_{\rm A}\,\sigma_{\rm A}^2$ as the functions of measured values $\omega$, $S\,\sigma$, and $\kappa\,\sigma^2$, and small parameter $\alpha$:
 \begin{empheq}[left = \empheqlbrace]{align}
 \label{wA}
 \omega_{\rm A} &~\simeq~ \omega~-~\alpha~\omega~,
 \\ \label{SsigA}
 S_{\rm A}\,\sigma_{\rm A} &~\simeq~ S\,\sigma~+~\alpha~(S\,\sigma-3\,\omega_{\rm A})~,
 \\ \label{ks2A}
 \kappa_{\rm A}\,\sigma_{\rm A}^2 &~\simeq~   \kappa\,\sigma^2 ~+~ \alpha~\left( \kappa\,\sigma^2-3\,\omega^2_{\rm A}-4\,\omega_{\rm A}\,S_{\rm A}\,\sigma_{\rm A} \right)~,&& \alpha,~|\beta|,~|\gamma|,~\ll~1~.
 \end{empheq}
Let us again use the argument of the non-observation of large fluctuations, which can be attributed to phase transition from hadron matter to quark-gluon plasma, or to the QCD critical point. Then, we may assume that the source of fluctuations is mainly a result of an interplay of resonance decays and other non critical effects, which, nevertheless, have to be understood and filtered out. 
The non-critical effects can be estimated assuming that the produced system is described by a gas of hadrons and resonances (HRG)~\cite{Karsch:2010ck,Garg:2013ata,Alba:2014eba}, or by relativistic mean-field nuclear
matter \cite{Fukushima:2014lfa}.
In ideal HRG with vanishing electric and strange chemical potentials, $\mu_Q=\mu_S=0$, one has for net baryon number~\cite{Karsch:2010ck}:
 \eq{\label{tanh}
 \frac{S_{\rm A}\,\sigma_{\rm A}}{\omega_{\rm A}} ~=~ \frac{\kappa_{\rm A}\,\sigma_{\rm A}^2}{\omega_{\rm A}^2} ~=~ \frac{1}{\omega_{\rm A}^2} ~=~ \tanh^2(\mu_B/T) ~\rightarrow~0~,~~~\mu_B/T\rightarrow0~,
 }
where $\mu_B$ is baryon chemical potential and $T$ is temperature of the created system. The ratio $\mu_B/T$ decreases fast with increasing energy of A+A reaction, therefore, $\tanh$ in Eq.~(\ref{tanh}) also decreases. The calculations of net proton number and net electric charge in HRG along the freeze-out line~\cite{Karsch:2010ck,Alba:2014eba} also show the decrease of ratios $S_{\rm A}\sigma_{\rm A}/\omega_{\rm A}$ and $\kappa_{\rm A}\sigma_{\rm A}^2/\omega_{\rm A}^2$. 
The introduction of interactions between particles in the form of either HRG with van der Waals interactions~\cite{Vovchenko:2016rkn}, or HRG with  excluded volume fitted to the Lattice data~\cite{Vovchenko:2017xad} further decreases these ratios, especially at low energies~\cite{Vovch:priv}.  
Fortunately, the increasing energy of A+A reactions also gives $\alpha,~|\beta|,~|\gamma|\ll 1$ for net charges, because small net charge gives even smaller $\langle n_{\rm A}\rangle$.
Therefore, one can introduce two more small parameters of the system,
\eq{\label{delta-epsilon}
 \delta~=~\frac{S_{\rm A}\,\sigma_{\rm A}}{\omega_{\rm A}}~,~~~\text{and}~~~\varepsilon~=~\frac{\kappa_{\rm A}\,\sigma_{\rm A}^2}{\omega_{\rm A}^2}~.
 }
Then, the system of equations (\ref{wIPM}-\ref{ks2IPM}) before any approximation looks as follows:
 \begin{empheq}[left = \empheqlbrace]{align}\label{MMCP-system}
 \omega &~=~ \omega_{\rm A}(1+\alpha)~, \nonumber \\
 S\,\sigma &~=~ \omega_{\rm A}~\frac{3\alpha+\delta+\alpha\,\beta}{1+\alpha}~, \nonumber \\
 \kappa\,\sigma^2 &~=~ \omega_{\rm A}^2~\frac{3\alpha+\varepsilon+\alpha\,\left[~6\beta+\gamma+4\delta~\right]}{1+\alpha}~.
  \end{empheq}
One can check that approximations $\alpha,~|\beta|,~|\gamma|,~\ll~1$ lead to (\ref{wIPM}-\ref{ks2IPM}). If, additionally, $3\alpha\ll\delta$ and $3\alpha\ll\varepsilon$, then $\omega\simeq\omega_{\rm A}$, $ S\,\sigma\simeq  S_{\rm A}\sigma_{\rm A}$, $ \kappa\,\sigma^2\simeq  \kappa_{\rm A}\sigma^2_{\rm A}$, and there is no need in any approximations. However, if this is not the case, but  $\alpha,~|\beta|,~|\gamma|,~\ll~1$, and $\delta\ll \varepsilon$, or  $\varepsilon\ll \delta$, then one can neglect either $\delta$ or $\varepsilon$, and solve Eqs.~ (\ref{wIPM}-\ref{ks2IPM}) or (\ref{MMCP-system}) numerically. In case of keeping $\varepsilon$ and neglecting $\delta$,
 \eq{\label{delta}
 |\delta|~\ll~|\varepsilon|~,
 }
one obtains
 \begin{empheq}[left = \empheqlbrace]{align}
 \label{wIPM-1}
 \omega &~=~ \omega_{\rm A}~(1+\alpha)~,
 \\ \label{SsigIPM-1}
  S\,\sigma 
  &~\simeq~ 3\,\alpha\,\omega_{\rm A}~,
  \\\label{ks2IPM-1}
  \kappa\,\sigma^2
   &~\simeq~ \kappa_{\rm A}\,\sigma_{\rm A}^2~(1-\alpha)
        ~+~3\,\alpha\,\omega_{\rm A}^2~, 
 && \alpha,~|\beta|,~|\gamma|~\ll~1~,~~~|\delta|~\ll~|\varepsilon|~,
 \end{empheq}
which has a simple analytical solution:
 \begin{empheq}[left = \empheqlbrace]{align}
 \alpha &~\simeq~ \frac{S\,\sigma}{3\,\omega-S\,\sigma}~,
 \\ \label{wks01}
 \omega_{\rm A} &~\simeq~ \omega~-~\frac{S\,\sigma}{3}~,
 \\ \label{wks02}
 \kappa_{\rm A}\,\sigma_{\rm A}^2 &~\simeq~  \frac{\omega_{\rm A}}{2\,\omega_{\rm A}-\omega}\left(\kappa\,\sigma^2 ~-~\omega_{\rm A}\,S\,\sigma\right)
 ~\simeq~ \kappa\,\sigma^2 ~-~\omega_{\rm A}\,S\,\sigma~,&& \alpha,~|\beta|,~|\gamma|~\ll~1~,~~~|\delta|~\ll~|\varepsilon|~.
 \end{empheq}
The solution can be also expressed through cumulants:
 \begin{empheq}[left = \empheqlbrace]{align}
 \alpha &~\simeq~  \left(\frac{ 3\,\kappa_2^2}{\langle N\rangle\,\kappa_3}~-~1\right)^{-1}
 \\  \label{wks1}
 \omega_{\rm A} &~=~\left(\frac{\kappa_2}{\langle N\rangle}\right)_{\rm A} ~\simeq~ \frac{\kappa_2}{\langle N\rangle}~-~\frac{1}{3}~\frac{\kappa_3}{\kappa_2} ~,
 \\
 \kappa_{\rm A}\,\sigma_{\rm A}^2 &~=~\left(\frac{\kappa_4}{\kappa_2}\right)_{\rm A} ~\simeq~ \frac{\kappa_4}{\kappa_2} ~-~\omega_{\rm A}~\frac{\kappa_3}{\kappa_2}~,&& \alpha,~|\beta|,~|\gamma|~\ll~1~,~~~|\delta|~\ll~|\varepsilon|~.
 \label{wks2}
 \end{empheq}
The approximate Eqs.~(\ref{wks01},\ref{wks02}) and (\ref{wks1},\ref{wks2}) remove fluctuations of participants and obtain the fluctuation of sources through {\it measured} values. This is the meaning of the MMCP method.

%
\section{Test of the MMCP in EPOS}\label{sec-EPOS}
%
%
The EPOS 1.99 model~\cite{Werner:2008zza,Pierog:2009zt} is a hadronic interaction model which does very well compared to experimental data at the SPS energy range~\cite{Aduszkiewicz:2015jna,Aduszkiewicz:2016mww}. In order to test relations introduced in Sec.~\ref{sec-Part} the large sample of $20\%$ most central Ar+Sc interactions at beam momentum $p_{lab}=150$~GeV/c is simulated. The centrality selection is based on the absolute impact parameter $b$, i.e. $20\%$ of events with the smallest $b$ were selected.
Fluctuations strongly depend on the analysis acceptance. In this paper the calculation are performed in the forward rapidity region assuming pion mass in center-of-mass frame ($y_{\pi}^{CMS}>0$) excluding the beam spectator domain~\cite{Grebieszkow:2007bv}, i.e. for each considered particle its $y^{CMS}_{p}$ (assuming $p$ mass) has to be smaller than $y^{*}_{beam}-0.5$.
For simplicity, detector efficiency is not considered.
The simulation test was performed for net electric charge defined as the difference between positively and negatively charged hadrons in the studied acceptance. The number of participants in a given event was obtained as a sum of variables $npj$ and $ntg$ in EPOS, which correspond to the number of primary projectile and target participants, respectively. The statistical uncertainty of scaled variance and higher order cumulant ratios were estimated using the bootstrap method~\cite{efron}.

It should be underlined that EPOS has much more complex way of producing particles than the simple wounded nucleon model. EPOS as a parton model describes the basis of high energy hadron-hadron interaction as an exchange of a 'parton ladder'. In this model, it contains two parts: the hard and a soft one~\cite{Pierog:2009zt}. In addition, there are two off-shell remnants (which behave as wounded nucleons in a limit of low energy) from projectile and target as well as core-corona effect for high density regions. Summarizing, in an event one can see up to three types of sources: nucleons, stings+nucleons and core+string+nucleons~\cite{Pierog:2009zt,Pierog:priv}.

\begin{figure}[h!]
\includegraphics[width=0.49\textwidth]{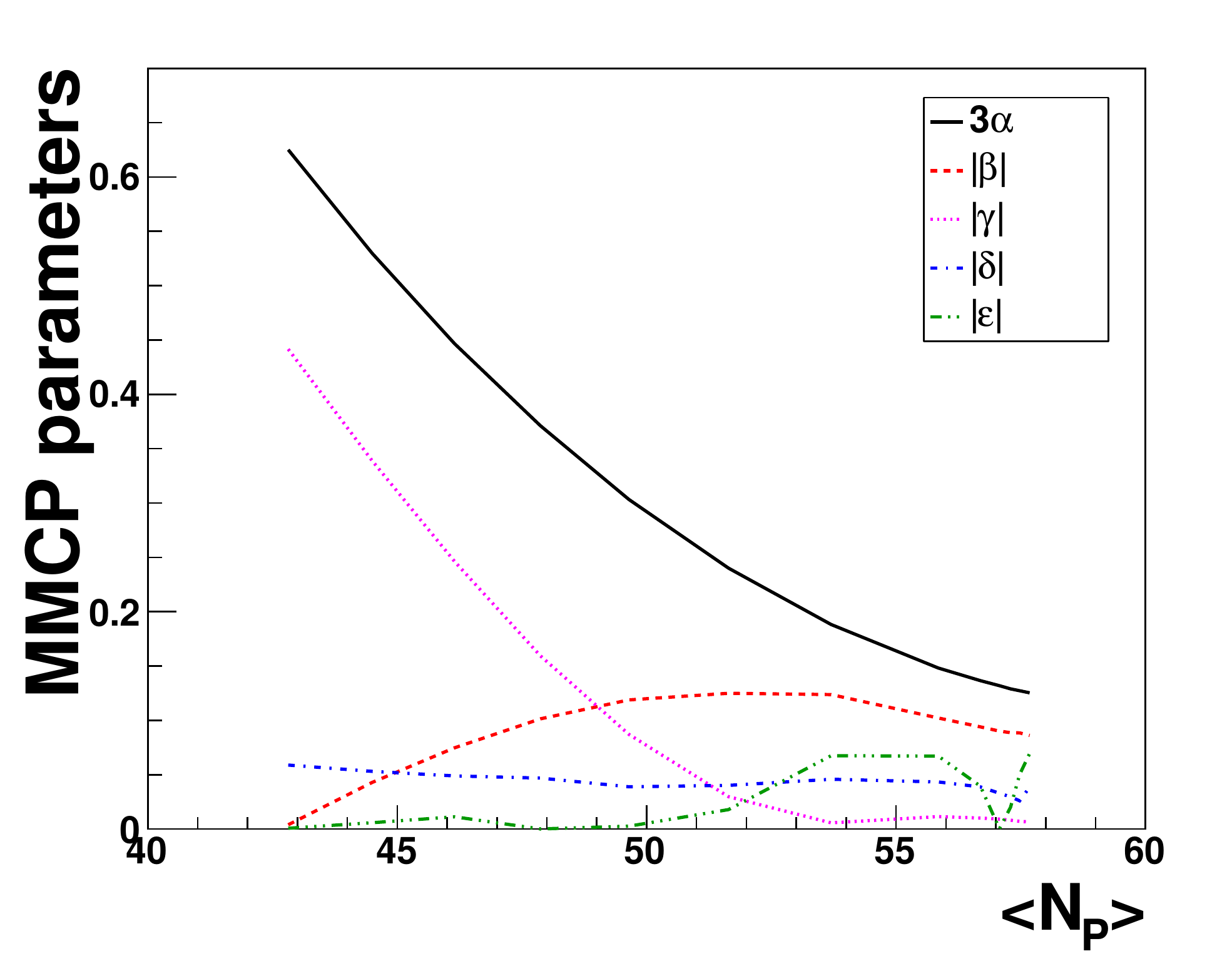}~~
\includegraphics[width=0.49\textwidth]{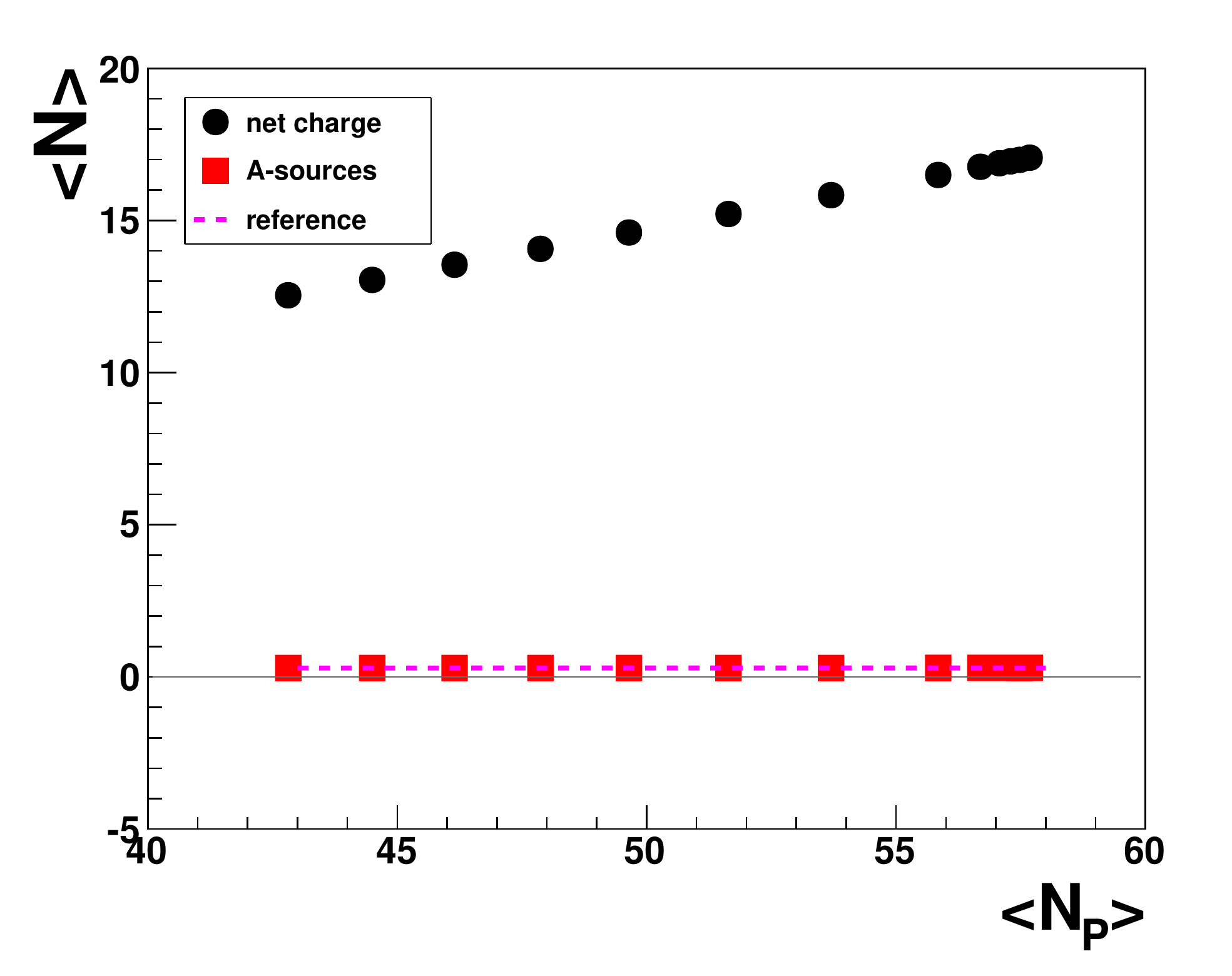}
\caption{Left: The MMCP parameters (\ref{alpha}), (\ref{beta-gamma}), (\ref{delta}) for different centrality windows, see text.
Right: The dependence of the net electric charge of the system on the average number of participants, $\langle N_{\rm P}\rangle$, in corresponding centrality windows (left to right): $20\%$, $17.5\%$, $15\%$, $12.5\%$, $10\%$, $7.5\%$, $5\%$, $2.5\%$, $1.5\%$, $1\%$, $0.75\%$, $0.5\%$ and $0.2\%$, with respect to zero centrality.
The label `net charge' corresponds to the values obtained in EPOS within the corresponding centralities, and not processed in any other way.
The net charge values produced by a source, $\langle n_{\rm A}\rangle$, are labelled as `A-source'.}\label{fig-0}
\end{figure}

\begin{figure}[h!]
\includegraphics[width=0.49\textwidth]{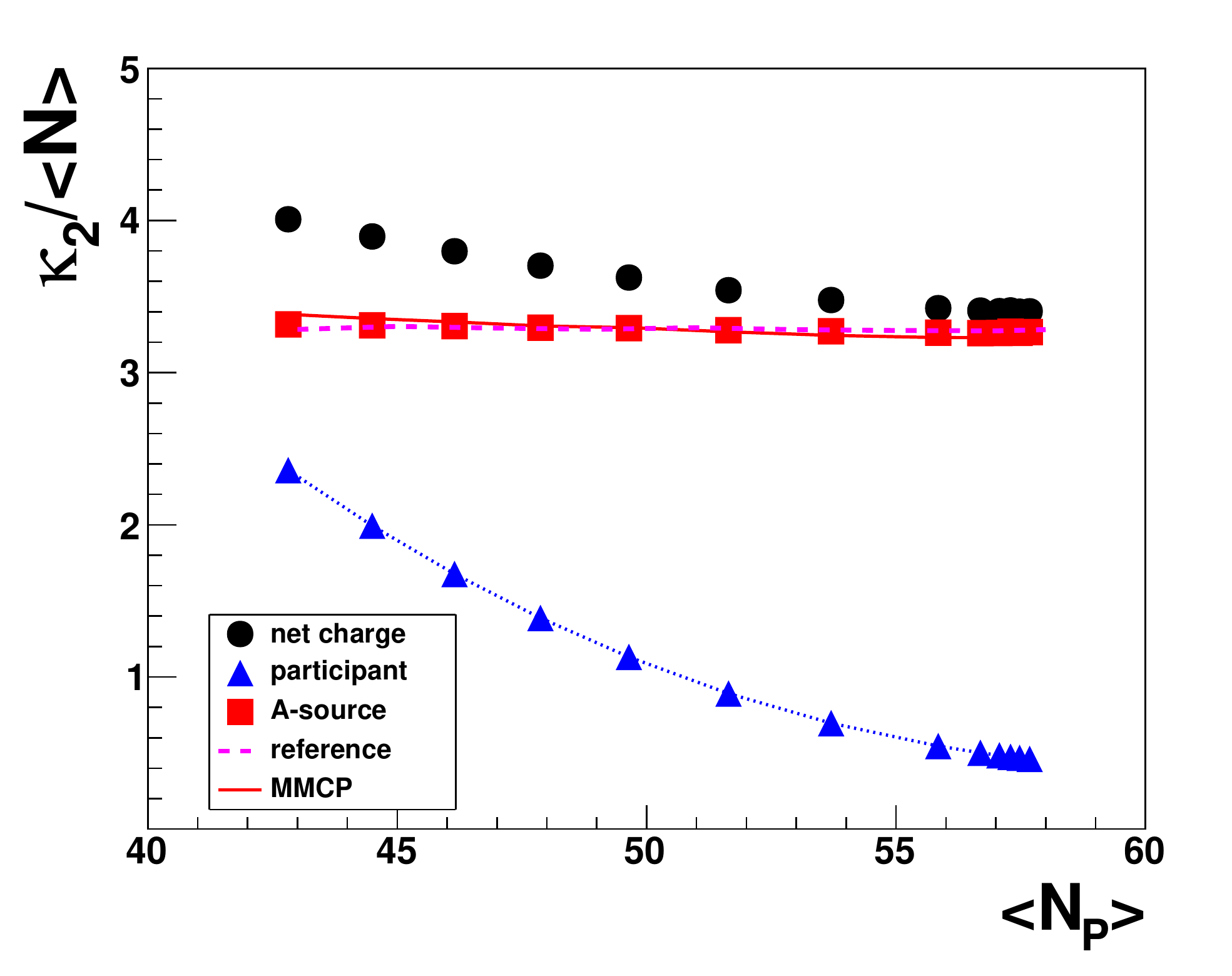}~~
\includegraphics[width=0.49\textwidth]{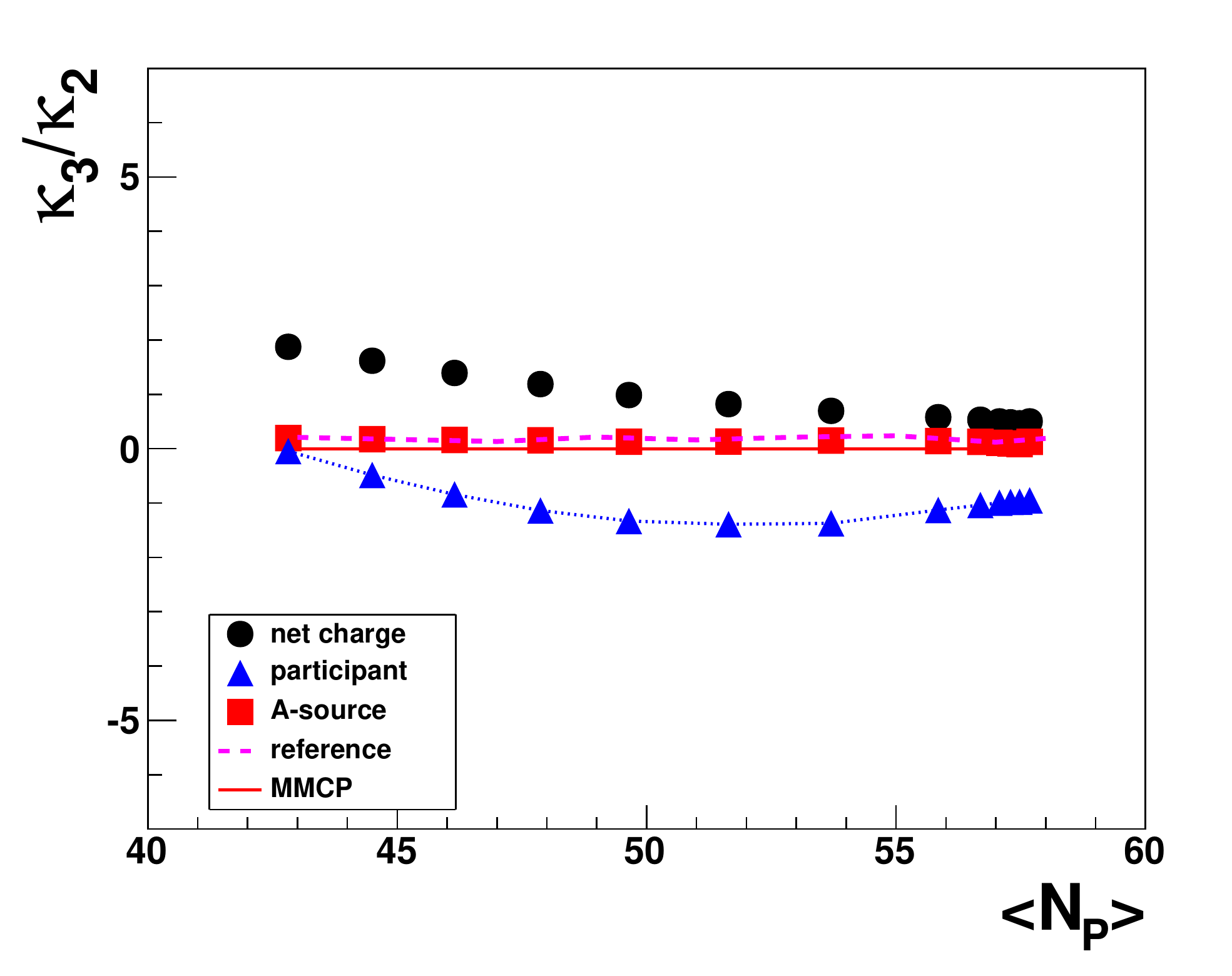}
\caption{Left: The same as in Fig.~\ref{fig-0} right for the scaled variance of the net electric charge, $\kappa_{2}/\langle N\rangle=\omega$, participant number, $\omega_{\rm P}$, scaled variance of a source, $\omega_{\rm A}$, and in the MMCP, see Eq.~(\ref{wks1}). The label `reference' corresponds to the values obtained for vanishing fluctuations of participants, see text. The solid line shows the MMCP results, see Eqs.~(\ref{wks1}) and (\ref{wks2}).
Right: The same for the normalized skewness, $\kappa_3/\kappa_2=S\,\sigma$. 
}\label{fig-1}
\end{figure}

\begin{figure}[h!]
\includegraphics[width=0.7\textwidth]{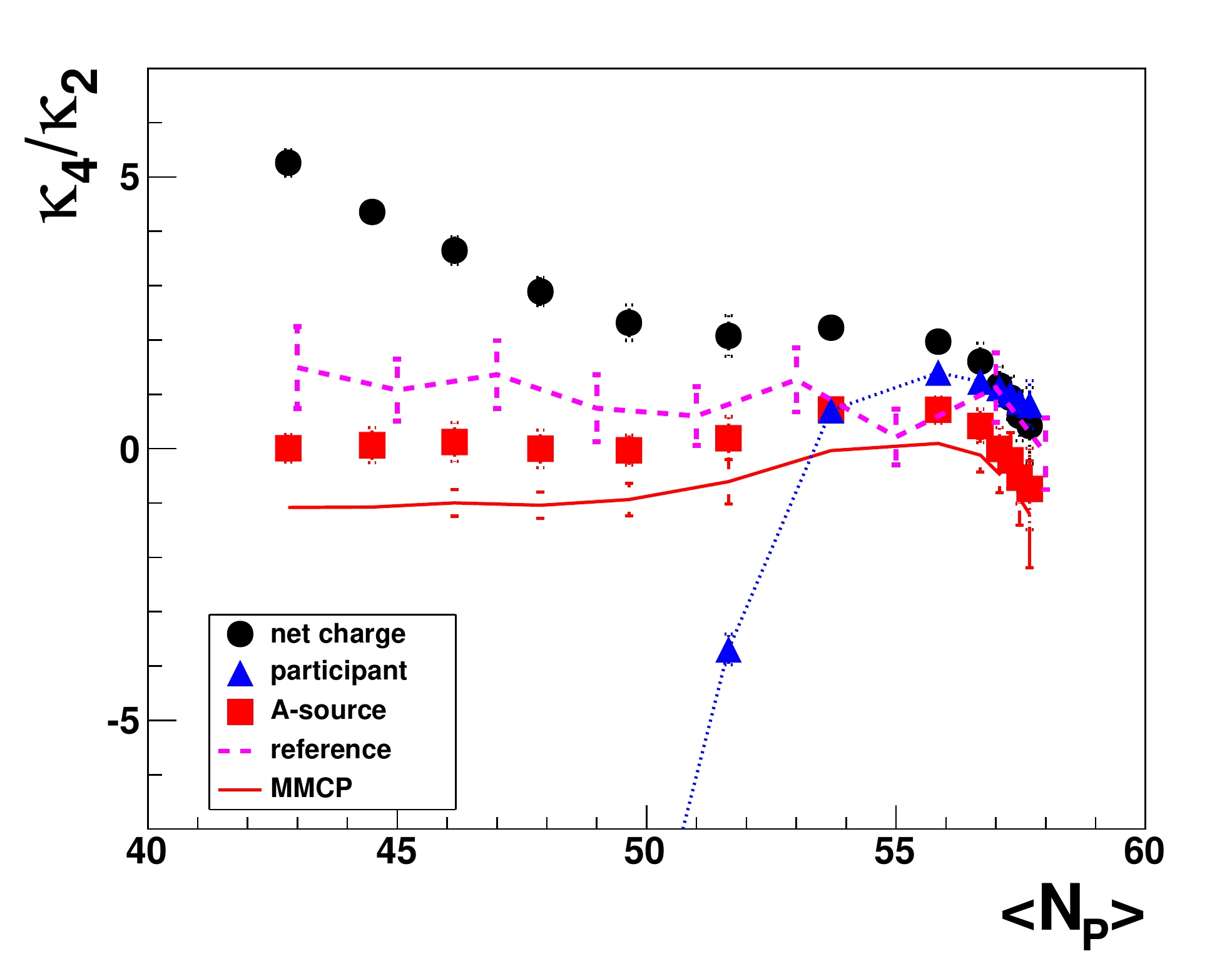}
\caption{The same as in Fig.~\ref{fig-1} for the normalized kurtosis, $\kappa_4/\kappa_2=\kappa\,\sigma^2$. }\label{fig-2}
\end{figure}

First of all, we check whether the assumptions of small participant fluctuations, $\alpha,~|\beta|,~|\gamma|~\ll~1$, are valid for the system that we study, see Fig.~\ref{fig-0} left. One can see that all the parameters are smaller than unity.
%
%
The $\alpha$ (\ref{alpha}) and $\gamma$ (\ref{beta-gamma}) decrease, when the bin width decreases (see caption to Fig.~\ref{fig-0} right), while $\beta$ increases. It means that the selection of more central Ar+Sc colisions decreases the relative importance of $\omega_{\rm P}$ and $\kappa_{\rm P}\sigma_{\rm P}^2$, but increases the normalized skewness of participant number distribution, $S_{\rm P}\sigma_{\rm P}$.  The $\delta$ (\ref{delta-epsilon}) is almost constant, which means that the centrality selection almost does not influence the relation between scaled variance and skewness of a source in the considered example, $S_{\rm A}\sigma_{\rm A}/\omega_{\rm A}$. It is interesting that $\varepsilon$ (\ref{delta-epsilon}), i.e. $\kappa_{\rm A}\sigma_{\rm A}/\omega_{\rm A}^2$ is sensitive to the bin width. The condition $|\delta|~\ll~|\varepsilon|$ is not always satisfied, however, $|\delta|,~|\varepsilon|<3\alpha$, therefore one can neglect both  $\delta$ and $\varepsilon$ in (\ref{MMCP-system}).
We check the approximate Eqs.~(\ref{wks1}) and (\ref{wks2}) 
and more general equations (\ref{Ssig}) and (\ref{ks2}), see Figs.~\ref{fig-1} and~\ref{fig-2}.
Each point corresponds to a given centrality window and the corresponding $N_{P}$ distribution. For example, the most left point corresponds to the centrality $c=0-20\%$ and $\langle N_{\rm P}\rangle=42.82(1)$, while the most right point corresponds to the centrality $c=0-0.2\%$ and $\langle N_{\rm P}\rangle=57.68(2)$.

The values labeled as `net charge' contain all possible sources of fluctuations of net electric charge, i.e. they include fluctuations of participants and fluctuations coming from a single source.
The participant number and its fluctuations are directly obtained from EPOS and are labeled as 'participant'.
The number of particles from a single source are defined as
 $\langle n_{\rm A}\rangle=
 \langle N\rangle/\langle N_{\rm P}\rangle=
 \langle N_{\rm net\, charge}\rangle/\langle N_{\rm participant}\rangle$.
The solid line shows the results of the MMCP method, which should be compared with square points coming from a single source and labeled as `A-source', calculated using Eqs.~(\ref{w}),~(\ref{Ssig}) and~(\ref{ks2}).

The `reference' line is obtained selecting $N_{P}=const$. In such case $\langle N_{P}\rangle=N_{P}$  and $\kappa_{2}=0$, meaning that $S_{P}\,\sigma_{P}$ and $\kappa_{P}\,\sigma_{P}^{2}$ are not defined, since they contain the division by $\kappa_2=0$ at this point. However, one can prove that the scaled variance, the normalized skewness, and the normalized kurtosis can be set to zero in this case.
The condition $N_{P}=const$ is equivalent to the replacement of the participant distribution, $P(N_{\rm P})$, by the uniform distribution, which is zero everywhere, except for the point $N_{P}=\langle N_{P}\rangle$.
For a uniform distribution that is non zero just between the points $N_{\rm P_1}$ and $N_{\rm P_2}$ the mean and higher central moments are~\cite{Mathworld}:
 \eq{
 \mu~=~\frac{N_{\rm P_1}+N_{\rm P_2}}{2}~, && \mu_2~=~\frac{(N_{\rm P_2}-N_{\rm P_1})^2}{12}~, && \mu_3~=~0~, && \mu_4~=~\frac{(N_{\rm P_2}-N_{\rm P_1})^4}{80}~,
 }
therefore,
 \eq{\label{Uniform}
 \omega_{\rm P}~=~\frac{1}{6}\,\frac{(N_{\rm P_2}-N_{\rm P_1})^2}{N_{\rm P_1}+N_{\rm P_2}}~, && S_{\rm P}\,\sigma_{\rm P}~=~0~, && \kappa_{\rm P}\,\sigma_{\rm P}^2~=~-\,\frac{(N_{\rm P_2}-N_{\rm P_1})^2}{10}~.
 }
One can see from (\ref{Uniform}) that for a narrow participant distribution, i.e. for $N_{\rm P_1}\rightarrow N_{\rm P_2}$, the scaled variance, the normalized skewness, and the normalized kurtosis vanish or are close to 0. Therefore, we can set $\omega_{\rm P}=S_{\rm P}\sigma_{P}=\kappa_{\rm P}\sigma_{\rm P}^{2}=0$ for the `reference' line.

The `A-source' and the `reference' fluctuations agree except for the normalized kurtosis. The difference arise due to the fact that `A-source' fluctuations are calculated for relatively wide centrality bins $\delta c$, while centrality is fixed, $\delta c=0$, for the `reference' line by definition. When the bin width decreases, $\delta c\rightarrow 0$, then the `A-source' and the `reference' agree within the uncertainty\footnote{The fact that the IPM equations (\ref{Nav}), (\ref{w}), (\ref{Ssig}), (\ref{ks2}) are valid for any number of participant, $N_{\rm P}$, while `net charge' and `A-source' fluctuations agree exactly, if number of participants does not fluctuate, means that EPOS 1.99 can be treated as an independent particle production model with respect to the net electric charge fluctuations in considered acceptance in spite of complicated internal dynamics. It means that Glauber Monte-Carlo fluctuations of participants dominate for the considered observables~\cite{Werner:priv}.}.
The same is the reason why the `x' coordinate of the `reference' is different from the `x' coordinate of the `net charge', `A-source' and `participant'. The latter are shown with the points that correspond to the average number of participants in the bin, $\langle N_{\rm P}\rangle$, while the `reference' can be calculated only for integer number of participants, and we selected $N_{\rm P}=43,~45,~47,~49,~51,~53,~55,~57,~58$. The hypothetical maximal number of participants in $^{40}_{18}$Ar~+~$^{45}_{21}$Sc reactions is $N_{\rm P}^{max}=40+45=85$. However, the participant number distribution drops so fast at large $N_{\rm P}$ that events with participant numbers $N_{\rm P}>60$ are rare.

The net electric charge is defined by the number of participating protons, which is roughly one half of the participant number for Ar+Sc, $\langle N_{\rm net~charge}\rangle/\langle N_{\rm P}^{max}\rangle = (18+21)/(40+45)\simeq 0.5$. The created particles are taken into account only if they have positive rapidity, which corresponds to one half of the created system\footnote{We do not take the whole system, because there is no net electric charge fluctuations in the whole rapidity range, because of global charge conservation.}. Therefore, the number of particles from one source (\ref{Nav}) should be $\langle n_{\rm A}\rangle=0.5*0.5=0.25$. The number obtained in this calculations is a bit larger, $\langle n_{\rm A}\rangle \simeq 0.3$, and, as a natural consequence of IPM, it is independent of centrality,  see Fig.~\ref{fig-0} right,
 \eq{
 \langle n_{\rm A}\rangle ~\simeq~ 0.3 ~\ll~ N_{\rm P}~,&&\text{for all}~\delta c~.
 }

In spite of so small number of particles from one source, $\langle n_{\rm A}\rangle$, it's fluctuations give the main contribution to the net charge fluctuations, which are also constant at all considered centrality windows
 \eq{
 \omega_{\rm A} ~\simeq~ 3.3 ~\sim~ \omega ~\gg~ \omega_{\rm P} \gtrsim 0.5~, && \text{for all}~\delta c~.
 }
Although fluctuations of participants are small, they exist, even if the centrality bin width approaches zero.
It means that if we choose to study total charge instead of the net charge in the same system, then we get larger contribution of fluctuations of participants.
\\

The skewness of a source is also independent on centrality and is close to zero, see Fig.~\ref{fig-1} right. It justifies the approximation~(\ref{delta}), which is necessary for the MMCP, and shows that the relatively large values of the net charge normalized skewness are due to the {\it second} moment fluctuations of participants,
 \eq{
 S_{\rm A}\,\sigma_{\rm A} ~\simeq~ 0.1~\ll~ S\,\sigma ~\simeq~ 3\,\langle n_{\rm A}\rangle\, \omega_{\rm P}~>~0~,&&\text{while}~S_{\rm P}\,\sigma_{\rm P}~\leq~0~,&&\text{for all}~\delta c~.
 }
The normalized skewness of participants, $S_{\rm P}\,\sigma_{\rm P}$, depends on $\langle N_{\rm P}\rangle$ and is negative for all considered centralities $c<20\%$. It gives a small contribution to the net charge skewness, because of the pre-factor $\langle n_{\rm A}\rangle^2\simeq 0.09$, see Eq.~(\ref{Ssig}). The negative skewness of the participant distribution means that it's mean value is on the left from the most probable value. This is in contrast to the results of~\cite{Braun-Munzinger:2016yjz}, which showed zero skewness of wounded nucleon distribution in $5\%-10\%$ most central collisions. The difference could, probably, be attributed to the fact that we use EPOS, while in~\cite{Braun-Munzinger:2016yjz} the authors use another model~\cite{Abelev:2013qoq}. Other possible reasons are that Pb+Pb reactions create a symmetric system, and Pb nucleus is heavy, thus, has small corona.
\\

The normalized kurtosis shows the bin width dependence for all types of fluctuations, see Fig.~\ref{fig-2}. For large centrality bins it is zero for a source, $\kappa_{\rm A}\,\sigma_{\rm A}^2\simeq 0$, while the fluctuations of participants are very large and negative, $\kappa_{\rm P}\,\sigma_{\rm P}^2\ll 0$. However, it almost does not influence net charge fluctuations, because the corresponding term is multiplied by $\langle n_{\rm A}\rangle^3\simeq 0.027$ in Eq.~(\ref{ks2}),
 \eq{
 \kappa_{\rm A}\,\sigma_{\rm A}^2 ~\simeq~ 0~,&&\kappa\,\sigma^2~\simeq~3\,\langle n_{\rm A}\rangle\,\omega_{\rm P}\,\omega_{\rm A}~>~0~,&&\kappa_{\rm P}\,\sigma_{\rm P}^2~\ll~0~,&&\text{for}~\delta c\geq 7.5\%~.
 }
The negative kurtosis of participants means that it's distribution is very flat at maximum.
For centrality windows smaller than $5\%$ the normalized kurtosises for all considered values slightly grow and then decrease to the values
 \eq{
  \kappa_{\rm A}\,\sigma_{\rm A}^2 ~\simeq~ -1~,&&\kappa\,\sigma^2~\simeq~\kappa_{\rm P}\,\sigma_{\rm P}^2~\simeq~1~,&&\text{for}~c=0-0.2\%~.
 }

The discrepancy between the `reference' and the MMCP in Fig.~\ref{fig-2} comes from the fact that $\beta$~(\ref{beta-gamma}) is not small enough in the considered example. Therefore, Eq.~(\ref{ks2-1}) reads
 \eq{\label{brn}
 \kappa\,\sigma^2 ~\simeq~ \kappa_{\rm A}\,\sigma_{\rm A}^2\,(1-\alpha)
        ~+~3\,\alpha\,\omega_{\rm A}^2 \left( 1~+~2\,\beta ~+~ ~\frac{1}{3}\,\gamma~ ~+~ \frac{4}{3}\,\delta  \right)~, 
 && \alpha~\ll~1~.
 }  
One can safely neglect $\gamma$ and $\delta$ in (\ref{brn}), but $2\beta$ can reach $-0.25$, see Fig.~\ref{fig-0}. Therefore, Eq.~(\ref{wks2}) requires the modification,
 \eq{
  \kappa_{\rm A}\,\sigma_{\rm A}^2 ~\simeq~ \frac{\kappa_4}{\kappa_2} ~-~\omega_{\rm A}~\frac{\kappa_3}{\kappa_2}\left( 1 ~+~ 2\,\beta \right)~,
 }
which gives the missing positive contribution, $-2\,\beta\,\omega_{\rm A}\,\kappa_3/\kappa_2$, to $\kappa_{\rm A}\,\sigma_{\rm A}^2$ in Fig.~\ref{fig-2}. However, the determination of $\beta$ requires the knowledge of $S_{\rm P}\,\sigma_{\rm P}$. The $\delta$ and $\varepsilon$ (\ref{delta-epsilon}) may also be too large to be neglected. Then, one needs more input in MMCP from a model, or from the measurements of higher fluctuation moments.

Our analysis shows that the higher the order of considered moments, the stronger is the dependence on the centrality bin width for both - the participant fluctuations, and fluctuations from a source.
Therefore, one needs to have a reliable estimation of these effects, in order to find the optimal bins and their widths, especially if one compares the results in different centrality bins (\ref{centbin}), or uses the CBWC procedure (\ref{cbwc}).

%
\section{Comparison of the MMCP and the CBWC methods}\label{sec-Comp}
%
%
The CBWC procedure~\cite{Luo:2011ts} means that a value $X$ is measured in $r$ sub-samples, and then summed up with the relative weights $w_r$ of the sub-samples $r$,
 \eq{
 X ~=~ \sum_r w_r~X_r~,\qquad w_r=n_{r}/\sum_r n_{r}~,
\label{CBWC_X}
 }
where $n_{r}$ is the number of events in the bin $r$.
The width of the sub-sample $r$ is chosen as small as possible, which is, for example, $\delta c=1\%$ in~\cite{Chen:2016xyu} and~\cite{Mukherjee:2016hrj}. Therefore, let say, $0-5\%$ centrality bin is further subdivided into 5 bins $0-1\%$, $1-2\%$, $2-3\%$, $3-4\%$, and $4-5\%$. The value $X$ is calculated in each sub-bin, and then summed up with $r$ running from 1 to 5.
We show that the CBWC procedure fails to remove participant fluctuations, if they are non-zero for an infinitely small sub-sample,
 \eq{\label{Pfluct}
 \omega_{{\rm P},r}~\rightarrow~\text{const}~=~\omega_{\rm P}^*~>~0~,\qquad\text{when}\qquad   \delta c_r~\rightarrow~ 0~.
 }
In order to explain the nature of the effect, we make a few simplifying assumptions.
We apply the CBWC to the scaled variance, normalized skewness and kurtosis, instead of their ingredients.
Suppose that we have the $0-5\%$ centrality bin, which we divide into 5 bins with equal multiplicity, so that $w_r=1/5$.
Assume that the number of sources and their fluctuations are constant in different sub-bins,
 \eq{
 \langle n_{{\rm A},r}\rangle ~=~\langle n_{\rm A}\rangle ~=~const ~,
\qquad \omega_{{\rm A},r} ~=~ \omega_{\rm A} ~=~const ~.
}
It is a realistic situation that is also realized in our Ar+Sc system, see Figs.~\ref{fig-0} and \ref{fig-1}. The number of particles produced by a source, $\langle n_{\rm A}\rangle$, and the scaled variance for the fluctuations from a source, $\omega_{\rm A}$, are independent on the bin width in the considered case.
Then the application of CBWC (\ref{CBWC_X}) to the scaled variance (\ref{w}) gives
 \eq{\label{CBWC}
 \omega_{_{\rm CBWC}}
 ~=~ \sum_{r=1}^5 \frac{1}{5}\,\omega_r ~=~ \omega_{\rm A}\sum_{r=1}^5\frac{1}{5} ~+~ \langle n_{\rm A}\rangle\sum_{r=1}^5 \frac{1}{5}\,\omega_{{\rm P},r}~.
 }
If participant fluctuations are also the same in different sub-bins,
 \eq{\label{wPconst}
 \omega_{{\rm P},r} ~\simeq~\omega_{\rm P}^* ~=~ const ~,
 }
then the CBWC just gives the same result as fluctuations in one sub-bin,
 \eq{\label{CBWC-1}
 \omega_{_{\rm CBWC}}
 ~\simeq~ \omega_{\rm A} ~+~ \langle n_{\rm A}\rangle ~ \omega_{\rm P}^*~.
 }
One can see that the fluctuations of participants, $\omega_{\rm P}^*$, are still there in (\ref{CBWC-1}).
The originally proposed CBWC is applied on the level of mean multiplicity, $\langle N\rangle$, and standard deviation, $\sigma$,  separately. This, however, does not change the qualitative effect,
 \eq{
 \sigma~=~\sqrt{\kappa_2}~=~\sqrt{\langle N\rangle\,\omega}~=~\sqrt{\langle N_{\rm P}\rangle\,\sigma_{\rm A}^2~+~\langle n_{\rm A}\rangle^2\,\sigma_{\rm P}^2}~,
 }
just the proof looks more complicated then (\ref{Pfluct}-\ref{CBWC-1}). One can also see from Fig.~\ref{fig-1} that there is no reason to select the centrality bin smaller than $\delta c=1\%$ for $\omega$ in the considered example, because fluctuations of participants saturate on the level about $\omega_{\rm P}=0.4$ for the five most right points around $\langle N_{\rm P}\rangle=57$, which correspond to $c=0-1.5\%$, $0-1\%$, $0-0.75\%$, $0-0.5\%$, $0-0.2\%$.

For the comparison of the MMCP and the original version of the CBWC the events from the $5\%$  most central collisions are divided into five approximately equal sub-bins: $0-1\%$, $1-2\%$, $2-3\%$, $3-4\%$ and $4-5\%$. Next, quantities $n_{r}$, $w_{r}\simeq 1/5 = 0.2$, $N$, $\sigma$, $S$ and $\kappa$ are obtained. Then the values $\sigma^2/\langle N\rangle=\omega=\kappa_2/\langle N\rangle$, $S\,\sigma=\kappa_3/\kappa_2$ and $\kappa\,\sigma^2=\kappa_4/\kappa_2$ are calculated for each sub-bin. Finally, the CBWC (\ref{CBWC_X}) is used to sum up the sub-bin values and obtain the values for the whole $0-5\%$ bin, see the last line in Table~\ref{Table1}.
\begin{table}[h!]
    \begin{center}
    \begin{tabular}{|c||c|c||c|c|c|c||c|c|c|}
        \hline
    ~bin width~ &   ~$n_{r}$~ &   ~$w_{r}$ &~   ~$\langle N\rangle$~   &   ~$\sigma$~    &   $S$ &   ~$\kappa$~ & ~$\sigma^2/\langle N\rangle$~ & ~$S\,\sigma$~ & ~$\kappa\,\sigma^2$~   \\
    \hline
    0-1\% & ~624827~ & ~0.198~ & ~16.88(1)~ & ~7.58(1)~ & ~0.0663(5)~ & ~0.020(1)~ & ~3.40(1)~ & ~0.503(4)~ & ~1.2(1)~ \\
    1-2\% & ~626043~ & ~0.199~ & ~16.36(1)~ & ~7.48(1)~ & ~0.0806(5)~ & ~0.043(1)~ & ~3.42(1)~ & ~0.603(4)~ & ~2.4(1)~ \\
    2-3\% & ~611242~ & ~0.194~ & ~15.83(1)~ & ~7.39(1)~ & ~0.0825(5)~ & ~0.039(1)~ & ~3.45(1)~ & ~0.610(4)~ & ~2.2(1)~ \\
    3-4\% & ~665988~ & ~0.211~ & ~15.32(1)~ & ~7.29(1)~ & ~0.0906(5)~ & ~0.055(1)~ & ~3.47(1)~ & ~0.660(4)~ & ~2.9(1)~ \\
    4-5\% & ~623110~ & ~0.198~ & ~14.81(1)~ & ~7.18(1)~ & ~0.1032(4)~ & ~0.044(1)~ & ~3.48(1)~ & ~0.742(3)~ & ~2.3(1)~ \\
    \hline
		0-5\% & ~3151210~ & ~1.0~ & ~15.833(4)~ & ~7.382(3)~ & ~0.0847(2)~ & ~0.0404(5)~ & ~3.442(3)~ & ~0.626(3)~ & ~2.20(3)~ \\
		\hline
    \end{tabular}
\end{center}
\caption{The first five rows correspond to the sub-bin values for the collected number of events $n_r$, the weight of the sub-bin $w_r$, the average net charge $\langle N\rangle$, standard deviation $\sigma$, skewness $S$, kurtosis $\kappa$ and their combinations. The last line corresponds to the values obtained for the whole $c\leq 5\%$ centrality using the standard CBWC~(\ref{CBWC_X}) for $\langle N\rangle$, $\sigma$, $S$, and $\kappa$. }\label{Table1}
\end{table}
The obtained CBWC values are compared with the fluctuations obtained without any processing - `net charge', with the fluctuations at fixed number of participants - `reference', with the fluctuations from a source - 'A source', and with the fluctuations obtained in the MMCP, see Table~\ref{Table2}.
\begin{table}[h!]
    \begin{center}
    \begin{tabular}{|c||c|c|c|c|c|}
        \hline
   ~$0-5\%$~    & ~CBWC~ & ~net charge~ & ~reference~ & ~A-source~  & ~MMCP~       \\
    \hline
    ~$\kappa_2/\langle N\rangle$~  & ~3.442(3)~ & ~3.477(3)~  & ~3.28(1)~ & ~3.27(1)~ & ~3.245(4)~  \\
    ~$\kappa_3/\kappa_2$~  	   & ~0.626(3)~ & ~0.697(1)~  & ~0.21(4)~ & ~0.15(6)~ & ~0~ 			  \\
    ~$\kappa_4/\kappa_2$~ 	   & ~2.20(3)~   & ~2.2(2)~     & ~1.3(5)~   & ~0.7(2)~   & ~0.0(2)~ \\
    \hline
    \end{tabular}
\end{center}
\caption{The comparison between fluctuation quantities, $\kappa_2/\langle N\rangle=\sigma^2/\langle N\rangle=\omega$, $\kappa_3/\kappa_2=S\,\sigma$, $\kappa_4/\kappa_2=\kappa\,\sigma^2$, obtained by different methods for the $c\leq 5\%$ centrality bin.
}\label{Table2}
\end{table}
The $0-5\%$ centrality corresponds to $\langle N_{P}\rangle=53.698(3)$ and to the seventh point, counting from the left, in all figures. The corresponding reference values are obtained for fixed $N_{P}=53$ from Eqs.~(\ref{w}), (\ref{Ssig}) and (\ref{ks2}). The fluctuations from a source are calculated using the same Eqs.~(\ref{w}), (\ref{Ssig}) and (\ref{ks2}), substituting the participant number and it's fluctuation moments from EPOS. The MMCP values are calculated using Eqs.~(\ref{wks1}) and (\ref{wks2}) from `net charge'.
In addition, calculations within the MMCP method for the widest considered centrality $c\leq 20\%$ are shown in
Table~\ref{Table3}.
\begin{table}[h!]
    \begin{center}
    \begin{tabular}{|c||c|c|c|}
        \hline
    ~$0-20\%$~  & ~net charge~ & ~A-source~    &   ~MMCP~    \\
    \hline
    ~$\kappa_2/\langle N\rangle$~  & ~4.008(5)~ & ~3.317(5) & ~3.383(6)~\\
    ~$\kappa_3/\kappa_2$~ & ~1.88(2)~ & ~0.20(8)~ & ~0~ \\
    ~$\kappa_4/\kappa_2$~ & ~5.3(2)~ & ~0.0(2)~ & ~-1.1(2)~ \\
    \hline
    \end{tabular}
\end{center}
\caption{The comparison between fluctuation quantities for the $c\leq 20\%$ centrality bin for `net charge', `A-source' and the MMCP only. The centrality bin corresponds to $\langle N_{P}\rangle=42.815(1)$, and to the most left point in Figs.~\ref{fig-1} and~\ref{fig-2}.}\label{Table3}
\end{table}

The CBWC procedure reduces statistical uncertainty, but overestimates scaled variance, normalized skewness and normalized kurtosis three times. In fact, the CBWC gives the average of net charge fluctuations over the selected sub-bins. One may notice that all the values in Table~\ref{Table1} monotonously change with the bin, except for $\kappa\,\sigma^2$ in the $0-1\%$ bin. Therefore, our selection of relatively equal sub-bins statistics gives $w_r\simeq0.2$, which leads to almost the same values in the final $0-5\%$ bin, and in the middle $2-3\%$ bin. A different selection of weights, $w_r$, would lead to a different results. Therefore, the CBWC depends also on the particular weights and bins selected for the CBWC.   
Although the difference is small for the scaled variance, the normalized skewness and kurtosis is overestimated by three times after the CBWC, compared `reference' or `A-source' in Table~\ref{Table2}. 

The MMCP method works good for the scaled variance. For the $0-5\%$ centrality the MMCP coincides within the uncertainty with the A-source values obtained using the complete knowledge of the participant fluctuations. For the $0-20\%$ centrality the scaled variance calculated in the MMCP deviates from the A-source only by two percents.
The normalized skewness of sources is zero by definition in the current version of the MMCP, see Eq.~(\ref{delta}). It agrees within three standard deviations with the `A-source' generated by EPOS. The normalized kurtosis in MMCP underestimates the `A-source' values, becoming closer to it, when the bin width decreases.
The MMCP formula for the normalized kurtosis~(\ref{wks2}) works better than fixing the bin width, but worse than Eq.~(\ref{wks1}) for the scaled variance $\omega=\kappa_2/\langle N\rangle$. This is the result of the fact that skewness of the participants, $S_{\rm P}\,\sigma_{\rm P}$, can not be neglected in  Eqs.~(\ref{wks1}) and (\ref{wks2}) and (\ref{delta}), as is assumed in Eq.~(\ref{beta-gamma}). Still, the obtained value is closer to the `A-source' than the `net charge'.
%

%
%
\section{Conclusions}\label{sec-Concl}
%
%
The MMCP (\ref{wks1}), (\ref{wks2}) works well for the considered example. The scaled variance calculated by the MMCP in the $0-5\%$ centrality bin coincides with the fluctuations from a source, as if there were no participant fluctuations. For the widest considered $0-20\%$ centrality the MMCP gives just 2\% error. It makes possible to use the MMCP almost independently of the event centrality selection.
It encourages us to look for the improvements of the MMCP by considering the moments higher then the current four. It should help to find better approximations for the third and the fourth moments of the source. 
It seems also to be interesting to test the current MMCP on symmetric systems like Pb+Pb, which could produce a symmetric distribution of participants, i.e. with zero skewness. We leave it for the future studies.

In the current example the average number of particles produced by a source, $\langle n_{\rm A}\rangle$, and it's fluctuations of the second, $\omega_{\rm A}$, and the third order, $S_{\rm A}\,\sigma_{\rm A}$, do not depend on the centrality bin width in the considered system.
However, the fourth order fluctuations of a source, $\kappa_{\rm A}\,\sigma_{\rm A}^2$, change non-monotonously for the bin width smaller than 5\% in the range from $-1$ to $+1$.
This change is clearly visible in all dependencies - `net charge', `A-source' and `reference'. Therefore, the effect should be taken into account for the fourth moment, and further studied for higher moments.

The $S\,\sigma$ and $\kappa\,\sigma^2$ depend on the lower order fluctuations, which give the largest contribution to their values. The fluctuations from a source that one would like to access, $S_{\rm A}\,\sigma_{\rm A}$ and $\kappa_{\rm A}\,\sigma_{\rm A}^2$, are almost zero:
 \eq{
 S\,\sigma ~\simeq~ 3\,\langle n_{\rm A}\rangle\, \omega_{\rm P}~>~0~, ~~~ \kappa\,\sigma^2~\simeq~3\,\langle n_{\rm A}\rangle\,\omega_{\rm P}\,\omega_{\rm A}~>~0~,
 &&
  S_{\rm A}\,\sigma_{\rm A} ~\simeq~0~,~~~ \kappa_{\rm A}\,\sigma_{\rm A}^2 ~\simeq~ 0~.
 }

The CBWC reduces statistical uncertainties, but is unable to remove participant fluctuations. It gives the average of the `net charge', i.e. non-processed fluctuations, which are dominated by participant fluctuations.
The result of the CBWC application depends on the width, weight, and the position of the sub-bins.  If a source is sensitive for the centrality selection, as it is the case for the normalized kurtosis $\kappa_{\rm A}\,\sigma_{\rm A}^2$ for the bin width smaller than 5\%, then the application of the CBWC may give arbitrary result.
The same is true if one does not mix the bins, but selects a particular centrality bin and increases statistics. The fluctuations of participants may persist even for the bin width that approaches zero.

\acknowledgments

The authors thank to W.~Broniowski, M.~Gazdzicki, M.~I.~Gorenstein, K.~Grebieszkow, T.~Pierog, and V.~Vovchenko
for fruitful discussions.
This work was partially supported by the National Science Center, Poland grant 2016/21/D/ST2/01983.

\bibliographystyle{h-physrev}
\bibliography{Npart20}

\end{document}